\documentclass[twocolumn,prX]{revtex4}
\usepackage{docs}

\usepackage{epsfig}

\usepackage{amssymb}

\newcommand{\bea}{\begin{eqnarray}}

\newcommand{\eea}{\end{eqnarray}}

\newcommand{\bean}{\begin{eqnarray*}}

\newcommand{\eean}{\end{eqnarray*}}

\newcommand{\nn}{\nonumber \\}

\newcommand{\be}{\begin{equation}}
\newcommand{\ee}{\end{equation}}

\def\spa#1.#2{\langle#1\,#2\rangle}
\def\spb#1.#2{[#1\,#2]}
\def\spab#1.#2.#3{\langle\mskip-1mu{#1} 
                  | #2 | {#3}]}

\def\spba#1.#2.#3{[\mskip-1mu{#1} 
                  | #2 | {#3}\rangle}

\def\spbb#1.#2.#3.#4{[\mskip-1mu{#1} 
                     | {#2} \ {#3} | {#4}]}

\def\spaa#1.#2.#3.#4{\langle\mskip-1mu{#1} 
                     | {#2} \ {#3} | {#4}\rangle}

\def\nn{{\nonumber}}

\newbox\SlashedBox
\def\slashed#1{\setbox\SlashedBox=\hbox{#1}
\hbox to 0pt{\hbox to 1\wd\SlashedBox{\hfil/\hfil}\hss}#1}
\def\hboxtosizeof#1#2{\setbox\SlashedBox=\hbox{#1}
\hbox to 1\wd\SlashedBox{#2}}

\newbox\charbox
\newbox\slabox
\def\s#1{{      
        \setbox\charbox=\hbox{$#1$}
        \setbox\slabox=\hbox{$/$}
        \dimen\charbox=\ht\slabox
        \advance\dimen\charbox by -\dp\slabox
        \advance\dimen\charbox by -\ht\charbox
        \advance\dimen\charbox by \dp\charbox
        \divide\dimen\charbox by 2
        \raise-\dimen\charbox\hbox to \wd\charbox{\hss/\hss}
        \llap{$#1$}
}}

\begin{document}
\begin{titlepage}

\title{Unitarity-Cuts and Berry's Phase}

\author{ Pierpaolo Mastrolia \\ 
   {\it Theory Group, Physics Department, CERN, 
        CH-1211 Geneva 23, Switzerland}
}

\begin{abstract}
Elaborating on the recent observation that
two-particle unitarity-cuts of scattering amplitudes can be 
computed by applying Stokes' Theorem,
we relate the Optical Theorem to the Berry Phase,
showing how the imaginary part of arbitrary one-loop Feynman amplitudes 
can be interpreted as the flux of a complex 2-form.
\end{abstract}

\maketitle

\thispagestyle{empty}
\end{titlepage}

Unitarity and geometric phases are two ubiquitous properties
of physical systems.

The Berry phase is the phase acquired by a system when it is subjected 
to a cyclic evolution, resulting only from the geometrical properties 
of the path traversed in the parameter space because of anholonomy
\cite{Berry,Shapere:1989kp}.

Unitarity represents the probability conservation in particle 
scattering processes described by the unitary {\it scattering
operator}, $S$. 
The relation, $S = 1 + i \ T$,
between the $S$-operator and the {\it transition operator}, $T$,
leads to the Optical Theorem,
\bea
 -i (T - T^{\dagger}) =  T^{\dagger} T \ .
\label{eq:basicOT}
\eea
The matrix elements of this equation between initial and final
states are expressed, in perturbation theory, 
in terms of Feynman diagrams. The evaluation of the right hand side requires
the insertion of a complete set of intermediate states.
Therefore, since $-i (T - T^{\dagger}) = 2 \ {\rm Im} T$,
Eq.(\ref{eq:basicOT}) yields the computation of the imaginary part of 
Feynman integrals from a sum of contributions from all possible
intermediate states. A Feynman diagram is thus responsible for an imaginary
part of the scattering amplitudes when the intermediate, virtual particles
go on-shell.

The Cutkosky-Veltman rules, implementing
the unitarity conditions, allow the calculation
of the discontinuity across a branch cut of an
arbitrary Feynman amplitude,
which corresponds to its imaginary part \cite{OldUnitarity}.
Accordingly, the imaginary
part of a given Feynman integral can be computed 
by evaluating the phase-space integral obtained 
by cutting two internal particles, which amounts 
to applying the on-shell conditions
and replacing their propagators by the corresponding 
$\delta$-function,
$(p^2 - m^2 + i0)^{-1} \to (2 \pi i) \ \delta^{(+)}(p^2 - m^2) .$

In later studies the problem of finding the discontinuity 
of a Feynman integral associated to a singularity was addressed in the language
of homology theory and differential forms \cite{homology}.

More recently multi-particle cuts have been combined 
with the use of complex momenta \cite{Britto:2004nc} for on-shell 
internal particles 
into very efficient techniques, by-now known as
unitarity-based methods,  
to compute scattering amplitudes for arbitrary 
processes - see \cite{Bern:2008ef,Bern:2007dw}
for a comprehensive list of references.

In this letter we establish an explicit relation between Unitarity 
and Berry's phase, by showing that the imaginary
part of a general {\it one-loop} Feynman amplitude, computed
by applying the Optical Theorem, can be interpreted as a Berry phase,
resulting from the curved geometry in effective momentum space 
experienced by the two on-shell particles
going around the loop.

In a recent work \cite{Mastrolia:2009dr}
it has been shown that
double-cuts of one-loop scattering amplitudes
can be efficiently evaluated by using the well-known
{\it Generalised Cauchy Formula}, also known as {\it Cauchy-Pompeiu Formula},
or {\it Cauchy-Green Formula} as well \cite{CauchyPompeiu}.
In the case of double-cuts, 
the 4-dimensional loop-momentum 
can be decomposed in terms of an {\it ad hoc} basis of four massless
vectors whose coefficients depend on two complex-conjugated 
variables, left over as free components after imposing the two 
on-shell cut-constraints.
Therefore, the double-cut phase-space integral 
is written as a two-fold integration over these two variables.
The integration is finally carried out by using 
Generalised Cauchy Formula as an application of Stokes' Theorem 
for rational function of two complex-conjugated 
variables.
As such, the result of the phase-space integration 
can be naturally interpreted as the flux of a 2-form that is given
by the product of the two tree-level amplitudes sewn along the cut.

\section{Double-Cut}
\label{sec:integration}

The two-particle Lorentz invariant phase-space (LIPS) in the $K^2$-channel
is defined as,
\bea
\int d^4\Phi \!\!&\equiv&\!\!\!\! 
\int d^4 \ell_1 
\ \delta^{(\!+\!)}(\ell_1^2 \!-\! m_1^2) 
\ \delta^{(\!+\!)}((\ell_1-K)^2 \!-\! m_2^2) \ , \quad
\label{def:phi4}
\eea
where $K^\mu$ is the total momentum across the cut.
We introduce a suitable parametrization
for $\ell_1^\mu$ \cite{Mastrolia:2009dr,ABFKM}, 
in terms of four massless momenta,
which is a solution of the two on-shell conditions,
$\ell_1^2 = m_1^2$ and $ (\ell_1-K)^2 = m_2^2$,
\bea
\ell_1^\mu = {1 - 2 \rho \over 1 + z \bar{z}} \Big(
p^\mu + 
z \bar{z} \ q^\mu
+  z \ \epsilon_{+}^\mu   
+  \bar{z} \ \epsilon_{-}^\mu \Big) + \rho K^\mu \ ,
\label{def:loopdeco}
\eea
where 
$p_\mu$ and $q_\mu$ 
are two massless momenta with the requirements,
\bea
&& \hspace*{-0.5cm}
p_\mu + q_\mu = K_\mu \ , \nn \\
&& \hspace*{-0.5cm}
p^2=q^2=0 \ , \quad 2 \ p\cdot q = 2 \ p\cdot K = 2 \ q\cdot K \equiv K^2 \ ;
\label{def:specialpq}
\eea
the vectors $\epsilon_{+}^\mu$ and 
$\epsilon_{-}^\mu$ 
are orthogonal to both
$p^\mu$ and $q^\mu$, with the following properties \footnote{
In terms of spinor variables 
that are associated to massless momenta, 
we can define
$p^\mu = (1/2){\spab p.\gamma^\mu.p }$ and 
$q^\mu = (1/2){\spab q.\gamma^\mu.q }$,
hence $\epsilon_{+}^\mu = (1/2){\spab q.\gamma^\mu.p }$
and $\epsilon_{-}^\mu = (1/2){\spab p.\gamma^\mu.q }$.},
\bea
\epsilon_{+}^2 = \epsilon_{-}^2 \!\! &=& \!\! 0 = 
\epsilon_{\pm}\cdot p = \epsilon_{\pm}\cdot q \ , \\
2 \ \epsilon_{+} \cdot \epsilon_{-} \!\! &=& - K^2 \ . 
\eea
The parameter $\rho$ is the pseudo-threshold,
\bea
\rho = {K^2 + m_1^2 - m_2^2 - \sqrt{\lambda(K^2, m_1^2, m_2^2)} 
       \over 2 K^2} \ ,
\eea
with the K\"allen function defined as,
\bea
\lambda(K^2, m_1^2, m_2^2) \!\! &=& \!\!  
(K^2)^2 + (m_1^2)^2 + (m_2^2)^2 \nn \\
\!\! & & \!\! 
- 2 K^2 m_1^2
- 2 K^2 m_2^2
- 2 m_1^2 m_2^2 \ , \quad
\eea
and depends only on the kinematics. \\
The complex conjugated variables $z$ and $\bar{z}$ parametrize
the degrees of freedom left over by the cut-conditions.

Analogously to the massless case \cite{Mastrolia:2009dr},
corresponding to the $\rho \to 0$ limit, 
because of (\ref{def:loopdeco}),
the LIPS in (\ref{def:phi4}) reduces to the remarkable expression,
\bea
\int d^4 \Phi = (1-2\rho)
\int \!\!\!\! \int
{dz \wedge d\bar{z} \over (1 + z \bar{z})^2} \ .
\label{eq:novelphi4}
\eea

\begin{figure}[t]
\begin{center}
\vspace*{-1cm}
\includegraphics[scale=0.15]{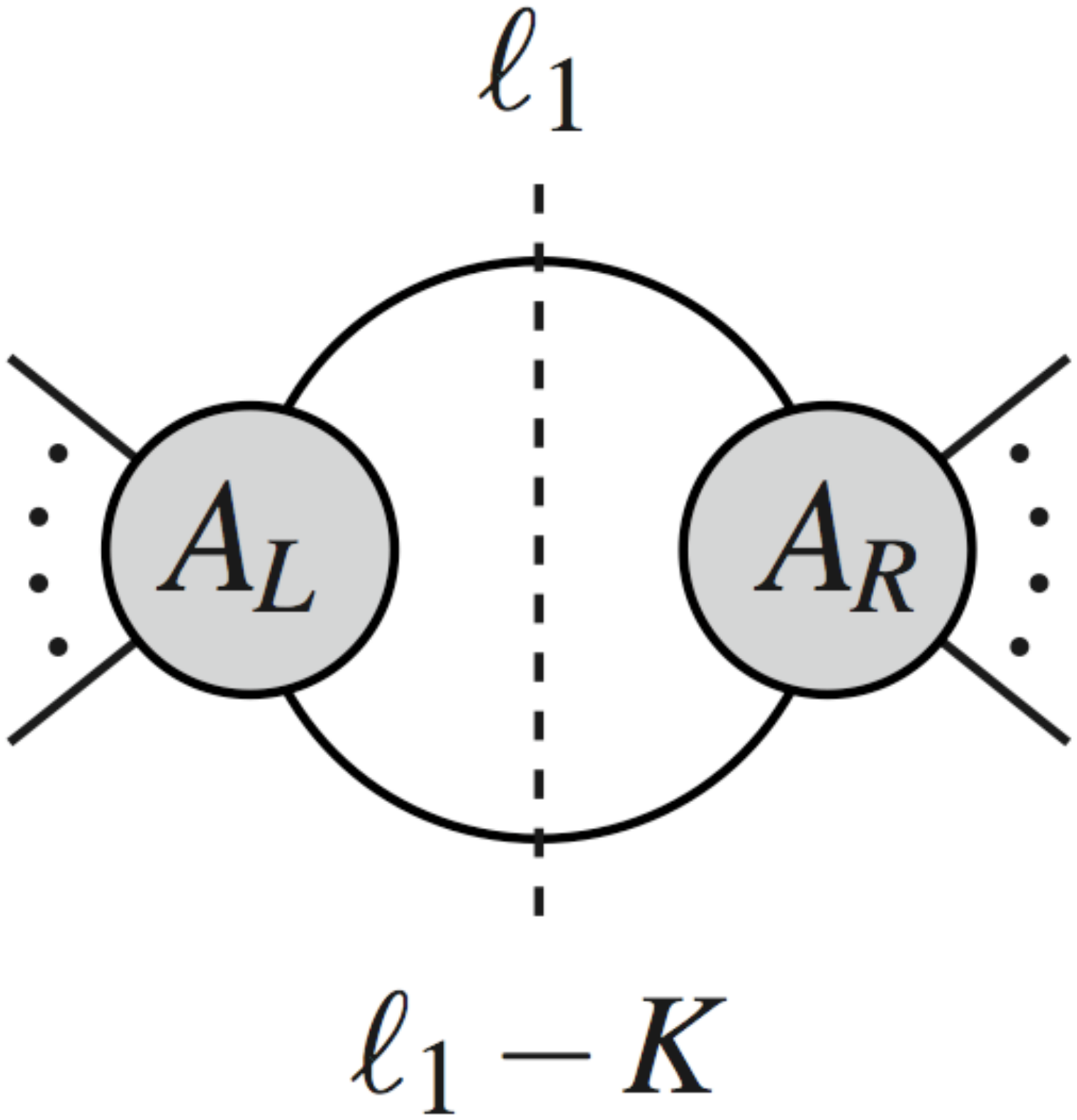}
\vspace*{-1.0cm}
\caption{Double-cut of one-loop amplitude in the $K^2$-channel.}
\label{fig:doublecut}
\end{center}
\end{figure}

The double-cut of a generic $n$-point amplitude
in the $K^2$-channel is defined as
\bea
\Delta \equiv 
\int d^4\Phi \  
A^{\rm tree}_L(\ell_1) \ 
A^{\rm tree}_R(\ell_1) \ ,
\eea
where $A^{\rm tree}_{L,R}$ are the tree-level amplitudes
sitting at the two sides of the cut, see Fig.\ref{fig:doublecut}.
By using (\ref{eq:novelphi4}) for the LIPS, 
and (\ref{def:loopdeco}) for the loop-momentum $\ell_1^\mu$,
one has,
\bea
\Delta =
(1-2\rho) \!\!
\int \!\!\!\! \int
dz \wedge d\bar{z} \ 
{ A^{\rm tree}_L(\rho, z, \bar{z}) \ 
A^{\rm tree}_R(\rho, z, \bar{z}) \over (1 + z \bar{z})^2} \ , \quad 
\label{def:zbarzInt}
\eea
where the tree-amplitudes $A^{\rm tree}_L$ and $A^{\rm tree}_R$ are rational
in $z$ and $\bar{z}$. Notice that $\rho$ is independent of $z$ and $\bar{z}$,
therefore its presence in the integrand does not affect the integration
algorithm. For ease of notation, we give the $\rho$-dependence
of the integrand as understood.

In \cite{Mastrolia:2009dr} we aimed at proposing an efficient 
method for computing the 
double-cut of one-loop scattering amplitudes. 
Accordingly, by applying a special version of the
so called {\it Generalised Cauchy Formula} 
also known as the {\it Cauchy-Pompeiu Formula} \cite{CauchyPompeiu}, 
one can write the two-fold integration 
in $z$- and $\bar{z}$-variables
appearing in Eq.(\ref{def:zbarzInt}) simply as 
a convolution of an unbounded $\bar{z}$-integral 
and a contour $z$-integral \footnote{
The roles of $z$ and $\bar{z}$ can be equivalently exchanged.
},
\bea
\Delta =
(1-2\rho) 
 \oint d z \!\! \int \!\! d\bar{z} \ 
{ A^{\rm tree}_L(z, \bar{z}) \ 
A^{\rm tree}_R(z, \bar{z}) \over (1 + z \bar{z})^2} \ 
 \ ,
\label{def:practicalzbarzInt}
\eea 
where the integration contour has to be chosen as enclosing all the 
complex $z$-poles.

In this letter we rather want to focus on 
what links Eq.(\ref{def:zbarzInt}) and Eq.(\ref{def:practicalzbarzInt}),
namely Stokes' Theorem \cite{Mastrolia:2009dr}, and on the geometrical
interpretation of its consequence:
the double-cut $\Delta$ in (\ref{def:zbarzInt}) is the flux of a 2-form.
It corresponds to an integral over the 
complex tangent bundle of the Riemann sphere,
where the curvature 2-form, $\Omega$, is defined as 
\footnote{
In \cite{Mastrolia:2009dr} it has been shown that the double-cut of the scalar
2-point function, $\Delta I_2$ = $\int d^4\Phi$ amounts to the integral 
$\int \!\!\! \int \Omega = - 2 \pi i$. This result corresponds
to the integration of the first Chern class, $(i/\pi) \int \!\!\! \int \Omega = 2$.
},
\bea
\Omega = 
{dz \wedge d\bar{z} \over (1 + |z|^2)^2} \ .
\eea
The product $A^{\rm tree}_L A^{\rm tree}_R $ is a rational function
of $z$ and $\bar{z}$, hence 
it can be written as ratio of two polynomials, $P$ and $Q$,
\bea
A^{\rm tree}_L(z, \bar{z}) \ 
A^{\rm tree}_R(z, \bar{z}) = {P(z,\bar{z}) \over Q(z,\bar{z})} \ ,
\eea
with the following relations among their degrees,
\bea
 {\rm deg}_z Q = {\rm deg}_z P \ , 
\qquad
 {\rm deg}_{\bar{z}} Q = {\rm deg}_{\bar{z}} P  \ .
\label{eq:degrees}
\eea

\section{Optical Theorem}

In the double-cut integral (\ref{def:zbarzInt}),
we did not make any assumptions on the tree-level amplitudes sewn along 
the cut, thus providing a general framework to the integration method developed in \cite{Mastrolia:2009dr}.
If we now choose $A_L^{\rm tree} = A_{m \to 2}^{*, \rm tree}$,
that is the conjugate scattering amplitude of a process $m \to 2$,
and $A_R^{\rm tree} = A_{n \to 2}^{\rm tree}$, that is the amplitude
of a process $n \to 2$, 
then $\Delta$ reads,
\bea
\Delta 
&=&
\int d^4\Phi \ 
 A^{*, \rm tree}_{m \to 2} \ A^{\rm tree}_{n \to 2}  
= \nn \\
\!\!&=&
- i \Big[ A^{\rm one-loop}_{n \to m} 
         - A^{*, \rm one-loop }_{m \to n}
    \Big] = \nn \\
\!\!&=&
2 \ {\rm Im}\Big\{ A^{\rm one-loop}_{n \to m} \Big\} \ ,
\label{eq:OpticalTheorem}
\eea
which is the definition of the 
two-particle discontinuity of the one-loop amplitude 
$A^{\rm one-loop}_{n \to m}$ across the branch cut in the  $K^2$-channel, 
corresponding to the field-theoretic version
of the Optical Theorem (\ref{eq:basicOT}) for one-loop Feynman amplitudes.

On the other side, because of Stokes' Theorem in (\ref{def:zbarzInt}, \ref{def:practicalzbarzInt}), one has,
\bea
\Delta 
\!\!&=&\!\!
(1-2\rho) 
\int \!\!\!\! \int \!\!
dz \wedge d\bar{z} \ 
{   A^{*, \rm tree}_{m \to 2} \ A^{\rm tree}_{n \to 2} 
\over (1 + z \bar{z})^2} \ 
= \nn \\
\!\!&=&\!\!
(1-2\rho) 
\oint dz \!\!
\int \!\! d\bar{z} \ 
{ A^{*, \rm tree}_{m \to 2} \ A^{\rm tree}_{n \to 2}
 \over (1 + z \bar{z})^2} \ ,
\label{eq:OpticalTheoremGeometric}
\eea
which provides a geometrical interpretation
of the imaginary part of one-loop scattering amplitudes,
as a flux of a complex 2-form through a surface bounded by the contour
of the $z$-integral (the contour should enclose all the poles in $z$ exposed
in the integrand after the integration in $\bar{z}$ \cite{Mastrolia:2009dr}).

Given the equivalence of (\ref{eq:OpticalTheorem}) and 
(\ref{eq:OpticalTheoremGeometric}), 
a correspondence between
the imaginary part of scattering amplitudes and the anholonomy of 
Berry's phase does emerge,
since the latter is indeed defined
as the flux of a 2-form in presence of curved 
space \cite{Berry,Shapere:1989kp}. 
In this context, one could establish a parallel description
between the Aharonov-B\"ohm (AB) effect and the double-cut of
one-loop Feynman integrals. 

In the AB-effect \cite{Aharonov:1959fk}, 
an electron-beam splits with half passing by 
either side of a long solenoid, before being recombined. 
Although the beams are kept away from the solenoid,  
so they encounter no magnetic field $({\bf B} = 0)$, they arrive 
at the recombination with a phase-difference 
that is proportional 
to the magnetic flux through a surface encircled by their paths.
The non-trivial anholonomy in this case is a consequence of Stokes' Theorem,
where the 2-form Berry curvature is written as the differential of 
the 1-form vector potential $(\nabla \times {\bf A})$. 

In the case of the double-cut of one-loop Feynman integrals,
we could describe the evolution of the system 
depicted in Fig.\ref{fig:doublecut}, from the left to the right.
The two particles produced in the $A_L$-scattering, going
around the loop and initiating the $A_R$-process, 
at the  $A_R$-interaction point would experience 
a phase-shift due to the non-trivial geometry in effective 
momentum space induced by the on-shell conditions.
As in the AB-effect, the anholonomy phase-shift is a consequence 
of Stokes' Theorem, and here it corresponds to the imaginary part of
the one-loop Feynman amplitude. \\ \\

\noindent
{\it -- Acknowledgements. }
I wish to thank Mario Argeri, Bruce Campbell, 
Gero von Gersdorf, Bryan Lynn, Ettore Remiddi 
and Aleksi Vuorinen,
for stimulating and clarifying discussions, and Michael Berry for 
his feedback on the manuscript.

\end{document}